\documentclass[12pt]{article}
\usepackage{amssymb}
\usepackage{amsmath,amsfonts,feynmp}
\usepackage{tipa}
\usepackage{stmaryrd}

\makeatletter\@addtoreset{equation}{section}\makeatother

\def\be{\begin{equation}}
\def\ee{\end{equation}}
\def\bea{\begin{eqnarray}}
\def\eea{\end{eqnarray}}

\makeatletter\@addtoreset{equation}{section}\makeatother

\renewcommand{\title}[1]{\vbox{\center\LARGE{#1}}\vspace{5mm}}
\renewcommand{\author}[1]{\vbox{\center#1}\vspace{5mm}}
\newcommand{\address}[1]{\vbox{\center\em#1}}

\begin{document}

\unitlength = .8mm

\begin{titlepage}
\begin{center}
\hfill \\
\hfill \\
\vskip 1.5cm

\title{ Consistent Boundary Conditions \\for New Massive Gravity in $AdS_3$ }

\vskip 0.5cm
 {Yan Liu\footnote{Email: liuyan@itp.ac.cn}} and {Ya-Wen
Sun\footnote{Email: sunyw@itp.ac.cn}}

\address{ Key Laboratory of Frontiers in Theoretical Physics£¬
\\ Institute of Theoretical Physics, Chinese Academy of Sciences
\\P.O. Box 2735, Beijing 100190, China}

\end{center}

\vskip 1cm

\abstract{ In this note we study the boundary conditions for the new
massive gravity theory in asymptotically $AdS_3$ spacetime. We find
that the Brown-Henneaux boundary conditions  are consistent with the
new massive gravity for all any value of the mass parameter.  At the
critical point where the central charge vanishes, the conserved
charges vanish, too. This provides further evidence that the theory
may be trivial at the critical point under Brown-Henneaux boundary
conditions.  The log boundary conditions are also examined and we
find that we can have
 three
kinds of log boundary conditions for this new massive gravity
theory, each of which could be consistent at the critical point,
while for other value of the mass parameter, the log gravity
boundary condition is not consistent. }

\vfill

\end{titlepage}


\section{Introduction}

Recently in \cite{Bergshoeff:2009hq} a new kind of massive gravity
 has been discovered in three dimensions. In this new massive
gravity, higher derivative terms are added to the Einstein Hilbert
action and unlike in topological massive gravity, parity is
 preserved in this new massive gravity. This new massive gravity is
 equivalent to the Pauli-Fierz action for a massive spin-2 field at
the linearized level in asymptotically Minkowski spacetime. In
\cite{Nakasone:2009bn, {Nakasone:2009vt}}, the unitarity of this new
massive gravity and the new massive gravity with a Pauli-Fierz mass
term was examined. Warped AdS black hole solutions for this new
massive gravity with a negative cosmological constant have been
found in \cite{Clement:2009gq}.

In \cite{Liu:2009bk}, the linearized gravitational excitations of
this new massive gravity around asymptotically $AdS_3$ spacetime has
been studied. There are four branches of highest weight graviton
solutions satisfying the Brown-Henneaux boundary conditions in this
theory: the left and right moving massless gravitons and the left
and right moving massive gravitons. It was found that there is also
a critical point for the mass parameter at which massive gravitons
become massless as in topological massive gravity
\cite{{Deser:1981wh},Deser:1982vy} in $AdS_3$ \cite{Li:2008dq}
(other interesting discussions on topological massive gravity theory
could be found in
\cite{Carlip:2008jk,{Hotta:2008yq},{Grumiller:2008qz},
{Li:2008yz},{Park:2008yy},{Grumiller:2008pr},{Carlip:2008eq},
{Gibbons:2008vi},{Anninos:2008fx},{Carlip:2008qh},{Giribet:2008bw},{Strominger:2008dp},
{Compere:2008cv},{Grumiller:2008es},{Garbarz:2008qn},{Blagojevic:2008bn},
{Henneaux:2009pw},{Maloney:2009ck}}). At the critical point, the
leftmoving and rightmoving central charges are both zero and the
energy of all branches of highest weight gravitons vanish under the
Brown-Henneaux boundary conditions. It was conjectured that the new
massive gravity may be trivial at the critical point under
Brown-Henneaux boundary conditions. However, the consistency of the
Brown-Henneaux boundary conditions with the new massive gravity in
asymptotically $AdS_3$ was not shown in \cite{Liu:2009bk}. In this
note we will first study the consistency of the Brown-Henneaux
boundary conditions with the new massive gravity in asymptotically
$AdS_3$. We find that Brown-Henneaux boundary conditions are
consistent with the theory for any value of the mass parameter. At
the critical point, both the left moving and right moving conserved
charges are zero. This provides further evidence that the new
massive gravity in $AdS_3$ may be trivial at the critical point
under Brown-Henneaux boundary condtions.

As is shown in \cite{Grumiller:2008qz}, at the chiral point of
topological massive gravity theory there can be a new kind of
solution which does not obey Brown-Henneaux boundary conditions.
This kind of solution appears at the chiral point because the left
moving massless graviton and the left moving massive graviton
degenerate at the point. To include this new solution to the theory,
the boundary conditions should be relaxed to the log boundary
conditions
\cite{{Grumiller:2008es},{Henneaux:2009pw},Maloney:2009ck}, and
under this kind of log boundary conditions in topological gravity in
$AdS_3$, the left moving conserved charge is no longer zero and the
theory is not chiral at the chiral point.  In this note, we show
that the new kind of solution also exists in the new massive
gravity. We can also take the log boundary conditions to include the
new solutions. We find that under the log boundary conditions, the
conserved charges are divergent for general value of the mass
parameter while at the critical point, the log boundary condition is
consistent with the theory.

Different from in Chiral gravity, because in this theory, the right
moving massless and massive gravitons also degenerate at the
critical point as well as the left moving modes, we can have two new
solutions. We can also relax the boundary conditions to the log
boundary condition in this theory to include the new solutions.
Depending on which solutions we want to include, we can have three
kinds of log boundary conditions: one for the left moving new
solution, one for the right moving and one for both modes. In the
first two cases, after imposing the log boundary conditions we can
only get nonzero conserved charge for one of the two modes and the
conserved charge for the other mode is still zero. This can be
viewed as a new kind of log chiral gravity, which is realized by
imposing different boundary conditions for the left moving and right
moving modes respectively. In the last case, we can have nonzero
conserved charges for both modes.

The structure of our work is as follows. In Sec.2 we will review the
new massive gravity in asymptotically $AdS_3$ and write out the
formula for the calculation of conserved charges in this theory. In
Sec.3 we examine the consistency of the Brown-Henneaux boundary
conditions. In Sec.4 we study the log boundary conditions. Sec.5 is
devoted to conclusions and discussions.

\section{The Basic Setup}
In this section we will first review the new massive gravity theory
\cite{Bergshoeff:2009hq,{Liu:2009bk}} with a negative cosmological
constant and then derive the useful formulae for calculating the
conserved charges with given boundary conditions.
\subsection{The New Massive Gravity Theory}
The action of the new massive gravity theory can be written
as\footnote{ We take the metric signature(-,+,+) and follow the
notation and conventions of MTW \cite{Misner:1974qy}. We assume
$m^2>0$ and $G$ is the three dimensional Newton constant which is
positive here.}\cite{Bergshoeff:2009hq}

 \be I=\frac{1}{16\pi
G}\int d^3x \sqrt{-g} \bigg[R-2\lambda-\frac{1}{m^2}K\bigg], \ee
where \be K=R^{\mu\nu}R_{\mu\nu}-\frac{3}{8}R^2,\ee $m$ is the mass
parameter of this massive gravity and $\lambda$ is a constant which
is different from the cosmological constant. The equation of motion
of this action is \be G_{\mu\nu}+\lambda
g_{\mu\nu}-\frac{1}{2m^2}K_{\mu\nu}=0 \label{eom}\ee where
 \be K_{\mu\nu}=-\frac{1}{2}\nabla^2 R
g_{\mu\nu}-\frac{1}{2}\nabla_\mu\nabla_\nu R+2\nabla^2 R_{\mu\nu}
+4R_{\mu \alpha \nu \beta}R^{\alpha
\beta}-\frac{3}{2}RR_{\mu\nu}-R_{\alpha\beta}R^{\alpha\beta}g_{\mu\nu}+\frac{3}{8}R^2g_{\mu\nu}.
\ee One special feature of this choice of $K$ is that $g^{\mu\nu}
K_{\mu\nu}=K.$

After introducing a non-zero $\lambda$, the new massive gravity
theory could have an $AdS_3$ solution
\begin{equation}\label{AdS3metric}
ds^2=\bar{g}_{\mu\nu}dx^\mu dx^\nu=\ell^2(-
\cosh^2{\rho}d\tau^2+\sinh^2{\rho}d\phi^2+d\rho^2),
\end{equation}
and the $\lambda$ in the action should be related to the
cosmological constant $\Lambda$ and the mass parameter by
 \be
m^2=\frac{\Lambda^2}{4(-\lambda+\Lambda)},\ee and \be
\Lambda=-1/\ell^2.\ee

It would be useful to introduce the light-cone coordinates
$\tau^{\pm}=\tau \pm \phi$, then the $AdS_3$ spacetime
(\ref{AdS3metric}) could be written as
\begin{equation}
ds^2=\frac{\ell^2}{4}(-d\tau^{+2}-2
\cosh{2\rho}d\tau^{+}d\tau^{-}-d\tau^{-2}+4d\rho^2).
\end{equation}
For convenience we define \be
{\cal{G}}_{\mu\nu}=R_{\mu\nu}-\frac{1}{2}g_{\mu\nu}R+\Lambda
g_{\mu\nu},\ee and by expanding
$g_{\mu\nu}=\bar{g}_{\mu\nu}+h_{\mu\nu}$ around $AdS_3$, we could
obtain the equation of motion for the linearized excitations
$h_{\mu\nu}$ as \be\label{eomold}
(2m^2+5\Lambda){\cal{G}}^{(1)}_{\mu\nu}-
\frac{1}{2}(\bar{g}_{\mu\nu}\bar{\nabla}^2-\bar{\nabla}_\mu\bar{\nabla}_\nu
+2\Lambda\bar{g}_{\mu\nu})R^{(1)}-2(\bar{\nabla}^2{\cal{G}}^{(1)}_{\mu\nu}-\Lambda\bar{g}_{\mu\nu}
R^{(1)})=0,\ee where
\begin{eqnarray} R_{\mu\nu}^{(1)}&
=& \frac{1}{2} (- \bar{\nabla}^2  {h}_{\mu\nu} - \bar{\nabla}_{\mu}
\bar{\nabla}_{\nu}  h + \bar{\nabla}^{\sigma} \bar{\nabla}_{\nu}
 h_{\sigma\mu} + \bar{\nabla}^{\sigma}  \bar{\nabla}_{\mu}  h_{\sigma\nu}),\\
R^{(1)} &\equiv& (R_{\mu\nu}  g^{\mu\nu})^{(1)} = - \bar{\nabla}^2 h
+ \bar{\nabla}_{\mu} \bar{\nabla}_{\nu}  h^{\mu\nu} - 2 \Lambda h,\\
{\cal{G}}_{\mu\nu}^{(1)}&=&R^{(1)}_{\mu\nu}-\frac{1}{2}\bar{g}_{\mu\nu}R^{(1)}-2\Lambda
h_{\mu\nu}.
\end{eqnarray}
After gauge fixing, we can solve the equation of motion
(\ref{eomold}) and obtain two sets of solutions. One set is the left
moving and right moving highest weight massless gravitons with
weights $(2,0)$ and $(0,2)$ respectively. The other set is the
$(\frac{6+ \sqrt{2+4m^2\ell^2}}{4},\frac{-2+
\sqrt{2+4m^2\ell^2}}{4})$ and $(\frac{-2+
\sqrt{2+4m^2\ell^2}}{4},\frac{6+ \sqrt{2+4m^2\ell^2}}{4})$ highest
weight massive gravitons. Both the non-negativeness of the central
charge and the non-negativeness of the mass of gravitons demand that
$m^2\ell^2\geq 1/2.$

Very analogous to  topological massive gravity theory in $AdS_3$,
interesting things happen at the special point $m^2\ell^2= 1/2.$ The
massive gravitons become massless at this point and both the central
charges and the energy of all branches of gravitons become zero.
This implies that the theory may be trivial at the special point
under Brown-Henneaux boundary conditions. In the next section we
will show that Brown-Henneaux boundary conditions are consistent
with this theory and at the critical point $m^2\ell^2= 1/2$ all
conserved charges vanish.

\subsection{Conserved Charges}
In this subsection we will give the basic formulae to calculate the
conserved charges using the covariant formalism
\cite{{Barnich:2001jy},Barnich:2007bf,{Guica:2008mu},{Maloney:2009ck}}
(see also
\cite{{abbottdeser},{iyerwald},{andersontorre},{torre},
{bbheneaux},{bbheneaux2},{barnichstokes},comperethesis})
 for this new massive gravity.

The covariant energy momentum tensor for the linearized
gravitational excitations of this new massive gravity theory can be
identified as \be\label{energy} 32 \pi m^2G
T_{\mu\nu}=(2m^2+5\Lambda){\cal{G}}^{(1)}_{\mu\nu}-
\frac{1}{2}(\bar{g}_{\mu\nu}\bar{\nabla}^2-\bar{\nabla}_\mu\bar{\nabla}_\nu
+2\Lambda\bar{g}_{\mu\nu})R^{(1)}-2(\bar{\nabla}^2{\cal{G}}^{(1)}_{\mu\nu}-\Lambda\bar{g}_{\mu\nu}
R^{(1)}).\ee It can be checked that the conservation of this energy
momentum tensor $\bar{\nabla}^{\mu}T_{\mu\nu}=0$ could be obtained
from the following equations: \bea\label{useful}
\bar{\nabla}^{\mu}{\cal{G}}^{(1)}_{\mu\nu}&=&0\nonumber\\
\bar{\nabla}^{\mu}({\bar{g}}_{\mu\nu}\bar{\nabla}^2 -
{\bar{\nabla}}_\mu {\bar{\nabla}}_\nu
+2\Lambda \bar{g}_{\mu \nu})R^{(1)} &=&0 \nonumber \\
\bar{\nabla}^{\mu}( \bar{\nabla}^2 {\cal{G}}^{(1)}_{\mu\nu} -
\Lambda  \bar{g}_{\mu \nu})R^{(1)}&=&0. \eea

It's shown in \cite{{Deser:2002rt},Deser:2002jk} that when the
background spacetime admits a Killing vector $\xi_\mu$, the current
\be {\cal{K}}^\mu=  16\pi G\xi_\nu T^{\mu\nu}\ee is covariantly
conserved $\bar{\nabla}_{\mu}{\cal{K}}^\mu=0$. Then there exists an
antisymmetric two form tensor ${\cal{F}}^{\mu\nu}$ such that
\be{\cal{K}}^\mu=16\pi G\bar{\nabla}_{\nu}{\cal{F}}^{\mu\nu}\ee and
the corresponding charge could be written as a surface integral as
\be \label{conserve}Q(\xi)=-\frac{1}{8\pi
G}\int_{M}\sqrt{-\bar{g}}{\cal{K}}^0 =-\frac{1}{8\pi
G}\int_{\partial{M}}dS_i\sqrt{{-\bar{g}}}{\cal{F}}^{0i},\ee where
$\partial{M}$ is the boundary of a spacelike surface $M$. We have
chosen $M$ as constant time surface here and the expression is under
the coordinate system of (\ref{AdS3metric}).

We can rewrite each term in the expression of the energy momentum
tensor (\ref{energy}) as a total covariant derivative term
\cite{{Deser:2002rt},Deser:2002jk} using the definition and 
the following properties of Killing vectors: \be\label{killingpro}
\bar{\nabla}_{\mu}\xi^\mu=0, \hskip 0.5 cm
\bar{\nabla}_{\sigma}\bar{\nabla}^{\mu}\xi^\sigma=2\Lambda \xi^\mu,
\hskip 0.5 cm \bar{\nabla}^2\xi_\sigma=-2\Lambda \xi_\sigma,\ee to
be \bea &&2\xi_\nu{\cal{G}}^{(1)\mu\nu}= \bar{\nabla}_\sigma \Big \{
\xi_\nu \bar{\nabla}^{\mu}h^{\sigma \nu} -\xi_\nu
\bar{\nabla}^{\sigma}h^{\mu\nu} +\xi^\mu \bar{\nabla}^\sigma h
-\xi^\sigma \bar{\nabla}^\mu h \nonumber \\
&& \hskip 2 cm + h^{\mu \nu}\bar{\nabla}^\sigma \xi_\nu - h^{\sigma
\nu}\bar{\nabla}^\mu \xi_\nu + \xi^\sigma \bar{\nabla}_{\nu}h^{\mu
\nu} -\xi^\mu \bar{\nabla}_{\nu}h^{\sigma \nu} +
h\bar{\nabla}^\mu \xi^\sigma \Big \} \\
&&\xi_\nu\Big({\bar{g}}^{\mu\nu}\bar{\nabla}^2 - {\bar{\nabla}}^\mu
{\bar{\nabla}}^\nu +2\Lambda g^{\mu
\nu}\Big)R^{(1)}=\bar{\nabla}_\sigma\Big\{\xi^\mu\bar{\nabla}^\sigma
R^{(1)}
+R^{(1)}\bar{\nabla}^\mu\xi^\sigma-\xi^\sigma\bar{\nabla}^\mu R^{(1)} \Big\}\\
&&\xi_\nu\Big( \bar{\nabla}^2 {\cal{G}}^{(1)\mu\nu} - \Lambda
\bar{g}^{\mu \nu}R^{(1)}\Big)=\bar{\nabla}_\sigma \Big \{\xi_\nu
\bar{\nabla}^{\sigma} {\cal{G}}^{(1)\mu \nu} - \xi_\nu
\bar{\nabla}^{\mu} {\cal{G}}^{(1)\sigma \nu} - {\cal{G}}^{(1)\mu\nu}
\bar{\nabla}^{\sigma} \xi_\nu \nonumber\\&&\hskip 5 cm +
{\cal{G}}^{(1)\sigma\nu}\bar{\nabla}^{\mu}\xi_\nu \Big
\}+2\Lambda\xi_\nu{\cal{G}}^{(1)\mu\nu}. \eea Thus we could write
\be\label{tmunu}  \xi_\nu
T^{\mu\nu}=\bar{\nabla}_{\sigma}{\cal{F}}^{\mu\sigma},\ee where \bea
{\cal{F}}^{\mu\sigma}&=&(1-\frac{1}{2m^2\ell^2})\frac{1}{2}\Big\{\xi_\nu
\bar{\nabla}^{\mu}h^{\sigma \nu} -\xi_\nu
\bar{\nabla}^{\sigma}h^{\mu\nu} +\xi^\mu \bar{\nabla}^\sigma h
-\xi^\sigma \bar{\nabla}^\mu h \nonumber \\
&&  + h^{\mu \nu}\bar{\nabla}^\sigma \xi_\nu - h^{\sigma
\nu}\bar{\nabla}^\mu \xi_\nu + \xi^\sigma \bar{\nabla}_{\nu}h^{\mu
\nu} -\xi^\mu \bar{\nabla}_{\nu}h^{\sigma \nu} + h\bar{\nabla}^\mu
\xi^\sigma\Big\}\nonumber\\&&-\frac{1}{4m^2}\Big\{\xi^\mu\bar{\nabla}^\sigma
R^{(1)}
+R^{(1)}\bar{\nabla}^\mu\xi^\sigma-\xi^\sigma\bar{\nabla}^\mu
R^{(1)}\Big\}\nonumber \\
&&  -\frac{1}{m^2}\Big\{\xi_\nu \bar{\nabla}^{\sigma}
{\cal{G}}^{(1)\mu \nu} - \xi_\nu \bar{\nabla}^{\mu}
{\cal{G}}^{(1)\sigma \nu} - {\cal{G}}^{(1)\mu\nu}
\bar{\nabla}^{\sigma} \xi_\nu +
{\cal{G}}^{(1)\sigma\nu}\bar{\nabla}^{\mu}\xi_\nu\Big\}.\eea Thus
the conserved charge (\ref{conserve}) becomes \be Q(\xi)
=-\frac{1}{8\pi
G}\int_{\partial{M}}dS_i\sqrt{{-\bar{g}}}{\cal{F}}^{0i}.\ee Here we
choose the spacelike surface as the constant time surface. For
asymptotic $AdS_3$ spacetime, the expression for the conserved
charge could be simplified as \be\label{concharge} Q(\xi)
=-\lim_{\rho \rightarrow\infty}\frac{1}{8\pi G}\int d\phi
\sqrt{{-\bar{g}}}{\cal{F}}^{0\rho},\ee where $\rho$ is the radial
coordinate of $AdS_3.$ Note that here to get the formula for the
conserved charges, we have used the definition of the Killing
vectors $\bar{\nabla}_{\mu}\xi_{\nu}+\bar{\nabla}_{\nu}\xi_{\mu}=0$
 which does not hold any more for asymptotic
symmetries of the spacetime. Thus for asymptotic symmetries which do
not obey
$\bar{\nabla}_{\mu}\xi_{\nu}+\bar{\nabla}_{\nu}\xi_{\mu}=0$,
(\ref{tmunu}) is no longer a conserved quantity and we need to add
some terms composed by
$\bar{\nabla}_{\mu}\xi_{\nu}+\bar{\nabla}_{\nu}\xi_{\mu}$ and
$h_{\mu\nu}$ \cite{Compere:2008cv}. However, the formula
(\ref{concharge}) will still be valid to the linearized level of
gravitational excitations in our consideration.

\section{Brown-Henneaux Boundary Condition}
In this section we will analyze the Brown-Henneaux boundary
 condition \cite{Brown:1986nw} for the new massive gravity theory. We
will calculate the conserved charges corresponding to the generators
of the asymptotical symmetry under this boundary condition and see
whether all the charges are finite or not.  In this and the
following sections we will work in the global coordinate system
(\ref{AdS3metric}).

The Brown-Henneaux boundary condition for the linearized
gravitational excitations in asymptotical $AdS_3$ spacetime can be
written as
  \be
\left(
  \begin{array}{ccccc}
 h_{++}= {\mathcal{O}}(1) & h_{+-}= {\mathcal{O}}(1)  & h_{+\rho}= {\mathcal{O}}({e^{-2\rho}})  \\
 h_{-+}=h_{+-} & h_{--}= {\mathcal{O}}(1)  & h_{- \rho}= {\mathcal{O}}({e^{-2\rho}})  \\
  h_{\rho+}=h_{+\rho} &
  h_{\rho-}=h_{- \rho} & h_{\rho\rho}= {\mathcal{O}}({e^{-2\rho}}) \\
  \end{array}
\right) \ee in the global coordinate system.

The corresponding asymptotic Killing vectors are
\bea\label{bhkilling}  \xi &=& \xi^+\partial_++\xi^-\partial_-+\xi^{\rho}\partial_\rho \nonumber\\
&=&~~[\epsilon^+({\tau^+})
+2{e^{-2\rho}}\partial_-^2\epsilon^-({\tau^-})+
{\mathcal{O}}({e^{-4\rho}})]\partial_+
\nonumber\\
&&+~ [\epsilon^-({\tau^-})
+2e^{-2\rho}\partial_+^2\epsilon^+({\tau^+})+
{\mathcal{O}}(e^{-4\rho})]\partial_- \nonumber\\&& -\frac{1}{2}
[\partial_{+}{\epsilon^{+}({\tau^+})} +\partial_-\epsilon^-(\tau^-)+
{\mathcal{O}}({e^{-2\rho}})]\partial_\rho. \eea

Because $\phi$ is periodic, we could choose the basis
$\epsilon_m^+=e^{im\tau^+}$ and $\epsilon_n^-=e^{in\tau^-}$and
denote the corresponding Killing vectors as $\xi^L_m$ and $\xi^R_n$.
The algebra structure of these vectors is \be\label{virasoro}
i[\xi^L_m, \xi^L_n]=(m-n)\xi^L_{m+n},~~i[\xi^R_m,
\xi^R_n]=(m-n)\xi^R_{m+n},~~[\xi^L_m, \xi^R_n]=0.\ee
 Thus
  these asymptotic Killing vectors give two copies
of Virasora algebra. To calculate the conserved charges using
(\ref{concharge}) we first parameterize the gravitons as follows
\bea\label{bhcon} h_{++}&=&f_{++}(\tau,\phi)+\dots\nonumber\\
h_{+-}&=&f_{+-}(\tau,\phi)+\dots\nonumber\\
h_{+\rho}&=&e^{-2\rho}f_{+\rho}(\tau,\phi)+\dots\nonumber\\
h_{--}&=&f_{--}(\tau,\phi)+\dots\nonumber\\
h_{-\rho}&=&e^{-2\rho}f_{-\rho}(\tau,\phi)+\dots\nonumber\\
h_{\rho\rho}&=&e^{-2\rho}f_{\rho\rho}(\tau,\phi)+\dots.,\eea where
$f_{\mu\nu}$ depends only on $\tau$ and $\phi$ while not on $\rho$
and the "$\dots$" terms are lower order terms which do not
contribute to the conserved charges. After plugging (\ref{bhcon})
into (\ref{concharge}) and performing the $\rho \rightarrow\infty$
limit, we obtain \be Q=\frac{1}{8\pi G\ell}\int d\phi
\Big[(1-\frac{1}{2m^2\ell^2})(\epsilon^+f_{++}+\epsilon^-f_{--})
-(1+\frac{1}{2m^2\ell^2})\frac{(\epsilon^++\epsilon^-)(16f_{+-}-f_{\rho\rho})}{16}\Big]\ee
for this theory. Three components of the equation of motion
(\ref{eomold}), which do not involve second derivative terms, can be
viewed as asymptotic constraints. The $\rho\rho$ component gives
 \be16f_{+-}-f_{\rho\rho}=0\ee at the boundary and the $+\rho$ and $-\rho$
 components give \be (1-2m^2\ell^2)\partial_-f_{++}=(1-2m^2\ell^2)\partial_+f_{--}=0\ee
 respectively. After plugging in these boundary constraints, the
 conserved charges become
\bea Q&=&\frac{1}{8\pi G\ell}\int d\phi
\big[(1-\frac{1}{2m^2\ell^2})(\epsilon^+f_{++}+\epsilon^-f_{--})\big]\nonumber\\~\nonumber\\&=&Q_{L}+Q_{R},
\eea where the left moving conserved charge is \be
Q_{L}=\frac{1}{8\pi G\ell}\int d\phi
\big[(1-\frac{1}{2m^2\ell^2})(\epsilon^+f_{++})\big],\ee and the
right moving conserved charge \be Q_{R}=\frac{1}{8\pi G\ell}\int
d\phi \big[(1-\frac{1}{2m^2\ell^2})(\epsilon^-f_{--})\big].\ee The
left moving and right moving conserved charges fulfill two copies of
Virasoro algebra with central charges \be
c_L=c_R=\frac{3\ell}{2G}\bigg(1-\frac{1}{2m^2\ell^2}\bigg),\ee which
are the same with the ones obtained in
\cite{{Liu:2009bk},{Kraus:2006wn},{Park:2006zw},{Park:2006fp},Park:2006pb}.
We can see that the conserved charges $Q$ are always finite for
arbitrary value of $m$, so the Brown-Henneaux boundary condition is
always consistent with the new massive gravity theory. At the
critical point $m^2\ell^2=1/2$, the conserved charges vanish which
provides further evidence to our previous conjecture
\cite{Liu:2009bk} that the new massive gravity theory may be trivial
at the critical point under the Brown-Henneaux boundary condition.

\section{Log Boundary Conditions}
At the critical point $m^2\ell^2=1/2$, new solutions of the equation
of motion (\ref{eomold}) can be constructed following
\cite{Grumiller:2008qz} to be
 \be{ \psi^{\rm
new}_{\mu\nu}\equiv\lim_{m^2\ell^2\to
1/2}\frac{\psi^M_{\mu\nu}(m^2\ell^2)-\psi^m_{\mu\nu}}{m^2\ell^2 -
1/2} }.\ee

Note here that we have both left and right moving massive and
massless modes, so we could obtain both left and right moving new
solutions, which are \be\label{leftlog} \psi^{\rm
newL}_{\mu\nu}\equiv\lim_{m^2\ell^2\to
1/2}\frac{\psi^{ML}_{\mu\nu}(m^2\ell^2)-\psi^{mL}_{\mu\nu}}{m^2\ell^2
- 1/2}=y(\tau,\rho)\psi^{mL}_{\mu\nu}  \ee and \be\label{rightlog}
\psi^{\rm newR}_{\mu\nu}\equiv\lim_{m^2\ell^2\to
1/2}\frac{\psi^{MR}_{\mu\nu}(m^2\ell^2)-\psi^{mR}_{\mu\nu}}{m^2\ell^2
- 1/2}=y(\tau,\rho)\psi^{mR}_{\mu\nu} \ee respectively. The function
\be y(\tau,\rho)=(-i\tau-\ln \cosh \rho)/2\ee is the same for the
two solutions.

These new solutions do not obey the Brown-Henneaux boundary
conditions. The energy of these new solutions can be calculated
using the Ostrogradsky procedure
\cite{Li:2008dq,{Grumiller:2008qz},{Buchbinder:1992pe}}  to be a
negative, finite and time-independent value, which is
$-\frac{49}{576G\ell^3}$ for the specific solutions (\ref{leftlog})
and (\ref{rightlog}). This may suggest the instability of the
$AdS_3$ vacuum under the relaxed boundary conditions. However, it
may still be stable non-perturbatively and it is still useful to
analyze the new boundary conditions.

\subsection{Left Moving Relaxation}

 In order to include these new interesting solutions, the
boundary conditions need to be loosened
\cite{{Grumiller:2008es},{Henneaux:2009pw},Maloney:2009ck}. Earlier
investigations on the relaxation of the boundary conditions for
 gravity coupled with scalar fields in anti-de Sitter spacetime
could be found in
\cite{{Henneaux:2002wm},{Henneaux:2004zi},{Hertog:2004dr},{Henneaux:2006hk},Amsel:2006uf}.
Because we have two new solutions, we can relax the boundary
condition either to include one of the two solutions or to include
both. Thus there are three kinds of boundary conditions. In this
subsection we analyze the first kind of these, which is exactly the
same as the one used in topological massive gravity for $\mu >0$.

We relax the boundary condition as follows\footnote{This boundary
condition is called the log boundary condition in the sense that if
we change from the global coordinate to the Poincare coordinate
system, the relaxed term is a logarithmic function of the radial
coordinate.} to include the solution $\psi^{\rm newL}_{\mu\nu}$:
  \be\label{boundary}
\left(
  \begin{array}{ccccc}
 h_{++}= {\mathcal{O}}(\rho) & h_{+-}= {\mathcal{O}}(1)  & h_{+\rho}= {\mathcal{O}}({
\rho e^{-2\rho}})  \\
 h_{-+}=h_{+-} & h_{--}= {\mathcal{O}}(1)  & h_{- \rho}= {\mathcal{O}}({e^{-2\rho}})  \\
  h_{\rho+}=h_{+\rho} &
  h_{\rho-}=h_{- \rho} & h_{\rho\rho}= {\mathcal{O}}({e^{-2\rho}}) \\
  \end{array}
\right). \ee Then the corresponding asymptotic Killing vector can be
calculated to be
\bea  \xi &=& \xi^+\partial_++\xi^-\partial_-+\xi^{\rho}\partial_\rho \nonumber\\
&=&~~[\epsilon^+({\tau^+})
+2{e^{-2\rho}}\partial_-^2\epsilon^-({\tau^-})+
{\mathcal{O}}({e^{-4\rho}})]\partial_+
\nonumber\\
&&+~ [\epsilon^-({\tau^-})
+2e^{-2\rho}\partial_+^2\epsilon^+({\tau^+})+ {\mathcal{O}}(\rho
e^{-4\rho})]\partial_- \nonumber\\&& -\frac{1}{2}
[\partial_{+}{\epsilon^{+}({\tau^+})} +\partial_-\epsilon^-(\tau^-)+
{\mathcal{O}}({e^{-2\rho}})]\partial_\rho. \eea Note that these
asymptotic Killing vectors are different from (\ref{bhkilling}) only
in the subleading order, so these also give two copies of Varasoro
algebra the same as (\ref{virasoro}).

With this new boundary condition we can parameterize the asymptotic
excitations as follows
\bea\label{logleft} h_{++}&=&\rho f^{L}_{++}(\tau,\phi)+\dots\nonumber\\
h_{+-}&=&
f^{L}_{+-}(\tau,\phi)+\dots\nonumber\\
h_{+\rho}&=&\rho e^{-2\rho}
f^{L}_{+\rho}(\tau,\phi)+\dots\nonumber\\
h_{--}&=&f^{L}_{--}(\tau,\phi)+\dots\nonumber\\
h_{-\rho}&=&e^{-2\rho}f^{L}_{-\rho}(\tau,\phi)+\dots\nonumber\\
h_{\rho\rho}&=&e^{-2\rho}f^{L}_{\rho\rho}(\tau,\phi)+\dots.\eea

Note that $f^{L}_{\mu\nu}$ depends only on $\tau$, $\phi$ while not
on $\rho$ and the ``$\dots$" terms are subleading terms which do not
contribute to the conserved charge. After plugging (\ref{logleft})
into (\ref{concharge}) and performing the $\rho \rightarrow\infty$
limit, we could obtain \be Q=\frac{1}{8\pi G\ell}\int d\phi
\bigg[(1-\frac{1}{2m^2\ell^2})(\infty)
-(1+\frac{1}{2m^2\ell^2})\frac{(\epsilon^++\epsilon^-)(16f^{L}_{+-}-f^{L}_{\rho\rho})}
{16}+\frac{2\epsilon^{+}f^{L}_{++}}{m^2\ell^2}\bigg].\ee

Here the first term is a linear divergent term proportional to
$\rho$ at infinity, which is caused by the relaxation of the
boundary condition. We see that the conserved charges can only be
finite at the special point $m^2\ell^2=1/2$, which means that the
log boundary condition is only well-defined at the special point
$m^2\ell^2=1/2$. There are now two asymptotic constraints coming
from the equation of motion (\ref{eomold}), which are \be
16f^{L}_{+-}-f^{L}_{\rho\rho}=0,\ee and \be \partial_-
f^{L}_{++}=0.\ee Now at the point $m^2\ell^2=1/2$, the conserved
charges become \be Q_L=\frac{1}{2\pi G\ell}\int d\phi
\big[\epsilon^+f^{L}_{++}\big], ~~~~ Q_{R}=0. \ee It's interesting
that though we have loosened the boundary condition to get nonzero
left moving charges, the right moving conserved charges are still
zero. The original new massive gravity theory has a left and right
moving symmetry and we have symmetric left and right moving modes.
However, the ``chiral" boundary condition (\ref{boundary}) breaks
this symmetry, which leads to a ``chiral" gravity which does not
possess the left-right symmetry anymore. Thus the new massive
gravity with the boundary condition (\ref{boundary}) can be viewed
as a new kind of log chiral gravity, the chirality of which is
realized by imposing ``chiral" boundary conditions. Note here that
the central charge for the left moving mode is still zero under this
boundary condition, similar to \cite{Henneaux:2009pw} for
topological massive gravity.

\subsection{Right Moving Relaxation}
The second kind of boundary condition is similar to the first one,
and can be obtained from the first one by exchanging $\tau^+$ and
$\tau^-$. This boundary condition is also the same to the log
boundary condition of chiral gravity with $\mu<0$.

We relax the boundary condition of the gravitons to be
 \be \left(
  \begin{array}{ccccc}
 h_{++}= {\mathcal{O}}(1) & h_{+-}= {\mathcal{O}}(1)  & h_{+\rho}= {\mathcal{O}}({e^{-2\rho}})  \\
 h_{-+}=h_{+-} & h_{--}= {\mathcal{O}}(\rho)  & h_{- \rho}= {\mathcal{O}}({\rho e^{-2\rho}})  \\
  h_{\rho+}=h_{+\rho} &
  h_{\rho-}=h_{- \rho} & h_{\rho\rho}= {\mathcal{O}}({e^{-2\rho}}) \\
  \end{array}
\right). \ee Then the corresponding asymptotic Killing vector is
\bea  \xi &=& \xi^+\partial_++\xi^-\partial_-+\xi^{\rho}\partial_\rho \nonumber\\
&=&~~[\epsilon^+({\tau^+})
+2{e^{-2\rho}}\partial_-^2\epsilon^-({\tau^-})+ {\mathcal{O}}({\rho
e^{-4\rho}})]\partial_+
\nonumber\\
&&+~ [\epsilon^-({\tau^-})
+2e^{-2\rho}\partial_+^2\epsilon^+({\tau^+})+
{\mathcal{O}}(e^{-4\rho})]\partial_- \nonumber\\&& -\frac{1}{2}
[\partial_{+}{\epsilon^{+}({\tau^+})} +\partial_-\epsilon^-(\tau^-)+
{\mathcal{O}}({e^{-2\rho}})]\partial_\rho. \eea These asymptotic
Killing vectors are different from (\ref{bhkilling}) only to the
subleading order, and it also gives two copies of Varasoro algebra
(\ref{virasoro}).

This time we can parameterize the asymptotic behavior of gravitons
as follows
\bea\label{logright} h_{++}&=& f^{R}_{++}(\tau,\phi)+\dots\nonumber\\
h_{+-}&=&
f^{R}_{+-}(\tau,\phi)+\dots\nonumber\\
h_{+\rho}&=&e^{-2\rho}
f^{R}_{+\rho}(\tau,\phi)+\dots\nonumber\\
h_{--}&=&\rho f^{R}_{--}(\tau,\phi)+\dots\nonumber\\
h_{-\rho}&=&\rho e^{-2\rho}f^{R}_{-\rho}(\tau,\phi)+\dots\nonumber\\
h_{\rho\rho}&=&e^{-2\rho}f^{R}_{\rho\rho}(\tau,\phi)+\dots..\eea

Note that $f^{R}_{\mu\nu}$ depends only on $\tau$, $\phi$ while not
on $\rho$ and the ``$\dots$" terms are lower order terms which do
not contribute to the conserved charges. After plugging
(\ref{logright}) into (\ref{concharge}) and performing the $\rho
\rightarrow\infty$ limit, we reach \be Q=\frac{1}{8\pi G\ell}\int
d\phi \bigg[(1-\frac{1}{2m^2\ell^2})(\infty)
-(1+\frac{1}{2m^2\ell^2})\frac{(\epsilon^++\epsilon^-)(16f^{R}_{+-}-f^{R}_{\rho\rho})}{16}+
\frac{2\epsilon^{-}f^{R}_{--}}{m^2\ell^2}\bigg].\ee

The first term is also a linear divergent term and we can see that
the conserved charges can only be finite at the critical point
$m^2\ell^2=1/2$, which means that the Log boundary condition is only
well-defined for this special point. The two asymptotical
constraints from the equation of motion is now \be
16f^{R}_{+-}-f^{R}_{\rho\rho}=0 \ee and \be \partial_+f^{R}_{--}=0.
\ee Thus at the critical point we have \be Q_{L}=0,~~~~
Q_R=\frac{1}{2\pi G\ell}\int d\phi \big[\epsilon^-f^{R}_{--}\big].
 \ee This is also a new chiral gravity with log boundary conditions.
 The right moving central charge is still zero.

\subsection{Both Modes Relaxation}

In this subsection we consider the possibility that both $\psi^{\rm
newL}$ and $\psi^{\rm newR}$ can be included after the boundary
conditions are relaxed. To reach this, we need to loosen the
boundary condition for the gravitational excitations to be
  \be
\left(
  \begin{array}{ccccc}
 h_{++}= {\mathcal{O}}(\rho) & h_{+-}= {\mathcal{O}}(1)  & h_{+\rho}= {\mathcal{O}}({\rho e^{-2\rho}})  \\
 h_{-+}=h_{+-} & h_{--}= {\mathcal{O}}(\rho)  & h_{- \rho}= {\mathcal{O}}({\rho e^{-2\rho}})  \\
  h_{\rho+}=h_{+\rho} &
  h_{\rho-}=h_{- \rho} & h_{\rho\rho}= {\mathcal{O}}({e^{-2\rho}}) \\
  \end{array}
\right). \ee The corresponding asymptotic Killing vector can be
calculated to be
\bea  \xi &=& \xi^+\partial_++\xi^-\partial_-+\xi^{\rho}\partial_\rho \nonumber\\
&=&~~[\epsilon^+({\tau^+})
+2{e^{-2\rho}}\partial_-^2\epsilon^-({\tau^-})+ {\mathcal{O}}({\rho
e^{-4\rho}})]\partial_+
\nonumber\\
&&+~ [\epsilon^-({\tau^-})
+2e^{-2\rho}\partial_+^2\epsilon^+({\tau^+})+ {\mathcal{O}}(\rho
e^{-4\rho})]\partial_- \nonumber\\&& -\frac{1}{2}
[\partial_{+}{\epsilon^{+}({\tau^+})} +\partial_-\epsilon^-(\tau^-)+
{\mathcal{O}}({e^{-2\rho}})]\partial_\rho. \eea Note that these
asymptotic Killing vectors are still different from
(\ref{bhkilling}) only to the subleading order, so it also gives two
copies of Varasoro algebra (\ref{virasoro}).

With this boundary condition we can parameterize the asymptotic
behaviors of gravitational excitations as follows
\bea\label{logdouble} h_{++}&=&\rho f^{B}_{++}(\tau,\phi)+\dots\nonumber\\
h_{+-}&=&
f^{B}_{+-}(\tau,\phi)+\dots\nonumber\\
h_{+\rho}&=&\rho e^{-2\rho}
f^{B}_{+\rho}(\tau,\phi)+\dots\nonumber\\
h_{--}&=&\rho f^{B}_{--}(\tau,\phi)+\dots\nonumber\\
h_{-\rho}&=&\rho e^{-2\rho}f^{B}_{-\rho}(\tau,\phi)+\dots\nonumber\\
h_{\rho\rho}&=&e^{-2\rho}f^{B}_{\rho\rho}(\tau,\phi)+\dots.\eea Here
$f^{B}_{\mu\nu}$ depends only on $\tau$, $\phi$ while not on $\rho$
and the ``$\dots$" terms are lower order terms which don't
contribute to the conserved charges. After plugging
(\ref{logdouble}) into (\ref{concharge}) and performing the $\rho
\rightarrow\infty$ limit, we have \bea Q&=&\frac{1}{8\pi G\ell}\int
d\phi \bigg[(1-\frac{1}{2m^2\ell^2})(\infty)
-(1+\frac{1}{2m^2\ell^2})\frac{(\epsilon^++\epsilon^-)(16f^{B}_{+-}-f^{B}_{\rho\rho})}{16}\nonumber\\
&&~~~~~~~~~+\frac{2\epsilon^{+}f^{B}_{++}+2\epsilon^{-}f^{B}_{--}}{m^2\ell^2}\bigg].\eea

The first divergent term is still linear divergent and the conserved
charges can only be finite at the special point $m^2\ell^2=1/2$.
This means that this third kind of log boundary condition is also
only well-defined at the special point. Different from the previous
two kinds of log boundary conditions, now the equation of motions
(\ref{eomold}) gives three asymptotic constraints, which are \be
16f^{B}_{+-}-f^{B}_{\rho\rho}=0\ee and \be
\partial_-f^B_{++}=\partial_+f^B_{--}=0 \ee respectively.

After drawing terms which vanish under these constraints, we get \be
Q_L=\frac{1}{2\pi G\ell}\int d\phi \big[\epsilon^+f^{B}_{++}\big],
~~ Q_{R}=\frac{1}{2\pi G\ell}\int d\phi
\big[\epsilon^-f^{B}_{--}\big] \ee at $m^2\ell^2=1/2$. This time the
left moving and right moving conserved charges are both nonzero. The
left and right moving central charges are still zero with the same
arguments in \cite{Henneaux:2009pw}.

\section{Conclusion and Discussion}
In this note we have studied the Brown-Henneaux and log boundary
conditions for new massive gravity in asymptotically $AdS_3$
spacetime. We find that the Brown-Henneaux boundary conditions are
always consistent with this theory and at a critical point
$m^2\ell^2=1/2$ the conserved charges all vanish and this can be
viewed as further evidence that the theory may become trivial at
this point under the Brown-Henneaux boundary conditions. Log
boundary conditions can also be imposed to this new massive gravity,
but it is only consistent at the critical point $m^2\ell^2=1/2$.
According to the boundary conditions of which components of the
gravitons are relaxed, we have three kinds of log boundary
conditions, and each of them is consistent at the critical point. It
is also interesting that the first two kinds of log boundary
conditions can give a new kind of log chiral gravity because we have
imposed boundary conditions which are not symmetric for the
coordinates $\tau^{+}$ and $\tau^{-}$.

Although the log boundary conditions for the new massive gravity can
be consistent to the linearized level, the physical consistency of
log gravity still needs to be studied. It will be interesting to
find solutions to the full equation of motion which can be viewed as
nonlinear generalizations of the linearized solutions which have the
log asymptotic behavior found in this paper, just as what has been
done for topological massive gravity in
\cite{AyonBeato:2004fq,{AyonBeato:2005qq}}. Also it would be
interesting to find other consistent boundary conditions for this
new massive gravity in asymptotically $AdS_3$ at both the critical
point and other values of the mass parameter. Further understanding
of this new massive gravity theory in asymptotically $AdS_3$ and the
dual field theory would be helpful to the study of quantum gravity
in three dimensions.

\section*{Acknowledgments}

We would like to thank Wei Song, Daniel Grumiller and Niklas
Johansson for helpful and valuable discussions. This work was
supported in part by the Chinese Academy of Sciences with Grant No.
KJCX3-SYW-N2 and the NSFC with Grant No. 10821504 and No. 10525060.

\end{document}